# From Centrality to Temporary Fame: Dynamic Centrality in Complex Networks

Dan Braha[1, 2] and Yaneer Bar-Yam[2]

[1] University of Massachusetts
Dartmouth, MA 02747, USA

[2] New England Complex Systems Institute
Cambridge, MA 02138, USA

We develop a new approach to the study of the dynamics of link utilization in complex networks using records of communication in a large social network. Counter to the perspective that nodes have particular roles, we find roles change dramatically from day to day. "Local hubs" have a power law degree distribution over time, with no characteristic degree value. Our results imply a significant reinterpretation of the concept of node centrality in complex networks, and among other conclusions suggest that interventions targeting hubs will have significantly less effect than previously thought.

Recent advances have demonstrated that the study of universal properties in physical systems may be extended to complex networks in biological and social systems [4, 5, 6, 10, 24]. This has opened the study of such networks to experimental and theoretical characterization of properties and mechanisms of formation. In this paper we extend the study of complex networks by considering the dynamics of the activity of network connections. Our analysis suggests that fundamentally new insights can be obtained from the dynamical behavior, including a dramatic time dependence of the role of nodes that is not apparent from static (time aggregated) analysis of node connectivity and network topology.

We study the communication between 57,158 e-mail users based on data sampled over a period of 113 days from log files maintained by the email server at a large university [18]. The time when an e-mail link is established between any pair of email addresses is routinely registered in a server, enabling the analysis of the temporal dynamics of the interactions within the network. To consider only emails that reflect the flow of valuable information, spam and bulk mailings were excluded using a prefilter. There were 447,543 messages exchanged by the users during 113 days observation. We report results obtained by treating the communications as an undirected network, where email addresses are regarded as nodes and two nodes are linked if there is an e-mail communication between them. Analysis based upon treating the network with asymmetric links (where a distinction is made between out-going links and incoming links) gave essentially equivalent results. From the temporal connectivity data, a time series of topological networks can be obtained; each represents an aggregation of links over a time scale that is short compared to the duration of observation (113 days). The edges forming each network in the time series thus represent the short time opportunity for communication as detected by the log files of the email server. Unless otherwise indicated, we set the time scale to one day, thus creating 113 consecutive daily networks.

Most studies of large social networks have accumulated data over the entire time of observation, whereas here using the smaller intervals of accumulation we can study how the network interactions change over time. Social network dynamics has historically been of interest, though data was limited [1,13]. Recent papers have considered the times between communications [14], the creation of temporally linked structures [15], or the addition of social links [16]. In this paper we study for the first time the dynamics of individual importance [21].

Our first result is that networks obtained on different days are substantially different from each other. Fig. 1 shows the correlation between corresponding edges of the 113 daily networks. Surprisingly, we find that all networks are weakly correlated, despite the expected routine nature of the social activity. Correlations between any two networks have a distribution that is approximately normal with a mean ± standard deviation of $0.15 \pm 0.05$ (we adopt this notation throughout). The low correlation implies that the existence of a link between two individuals at one time does not make it much more likely that the link will appear at another time. While all networks are weakly correlated, we find that workdays and weekends are more distinct, so that workday networks, and weekend networks are more correlated among themselves (correlations $0.17 \pm 0.03$ and $0.16 \pm 0.05$, respectively), than they are with each other (correlation $0.12 \pm 0.02$). Remarkably, the low correlations increase only very gradually if we form networks using data over multiple days, and never reach a high value even if networks are made from communications over a month or more [Fig. 1(b)].

Using the nodal "degree" (the number of nodes a particular node is connected to) we characterized the centrality of nodes in the daily networks. Each of the daily networks has a distribution of nodal degrees well described by a power-law [12], with exponents in the range $[2.5, 2.8]$. Thus a small number of highly connected nodes have great importance in the connectivity of the network. However, while each daily network has highly connected nodes, we found that they were not the same nodes. The degree of a node varied dramatically over time. For each identified 'local hub,' we measured its degree from day to day over the duration of observation. Surprisingly, we find that a large number of 'local hubs' exhibit a highly fluctuating time-series (Fig. 2). The corresponding distribution of degrees over time itself follows a scale-free power-law distribution over two orders of magnitude (Fig. 2). The degree distribution of a hub over time implies that the node's degree does not have a characteristic value. The degree is small most of the time, but we only need to wait long enough to encounter degrees of any size.

A broader characterization of which nodes are important on a given day was made by comparing how the nodes were ranked in importance. We identified the top 1000 nodes, about 1.7% of the network according to their degree, for each of the daily networks. We then determined, for each pair of daily networks, the percentage of nodes that appear in both top-ranking lists ("centrality overlap," Fig. 3). The centrality overlap between any two networks is small, around $0.27 \pm 0.06$. When considering separately workday and weekend networks, the overlap values are around $0.33 \pm 0.03$ and $0.20 \pm 0.04$, respectively; consistent with the bimodal nature of the social activity. The distinctiveness of the top 1,000 nodes between daily networks is also typical for other top-ranking list sizes. By varying the percentage of nodes in the top-ranking list, it is found that the mean centrality overlap, which is already small for small percentages (0.3), actually decreases to a value of 0.2 at around 4%, before increasing slowly to 1 when the list includes all the nodes. The distributions of ranking overlaps are well behaved, having a standard deviation much smaller than the mean.

We compared daily networks with the aggregate network, as would be considered by other contemporary studies, by aggregating over the entire 113 day observation. Our previous results suggest, and direct analysis confirms, that daily networks deviate significantly from the aggregate network. We determined which nodes in the daily 1000 top-ranking list also appear in the top-ranking list of the aggregate network, obtaining the binary image in Figure 4(a). Though some nodes that are ranked high in the daily networks are also ranked high in the aggregate network, a significant number are not. In particular, we find that the centrality overlap is $0.41 \pm 0.03$ and $0.27 \pm 0.04$, for weekday and weekends respectively. Comparing other sizes of the top ranked nodes gives similar results. Perhaps even more surprisingly, the nodes that are highly ranked in the aggregate network are not even *on-average* important in daily networks. To show this we calculated the average ranking position of the top 1000 highly connected nodes in the aggregate network for each daily network. The average ranking position over time (normalized to a fraction so that 1 is the highest and 0 is the lowest) exhibits a weekly oscillation from about 0.40 to 0.65. In the aggregate network these nodes have an average ranking of 0.99. This shows that highly connected nodes in the aggregate network only play a moderate role in the daily networks.

Finally, we considered a full range of networks formed by aggregating links over time scales that are longer than a day and shorter than the full time period [Fig. 4(b)]. Similar relationships between smaller and larger time scales to those found above are observed. Moreover, the similarity between networks at a particular time scale increases as a power-law, so there is no particular time scale at which a converged structure is achieved. Thus, the network dynamics follows a "multiscale" structure with networks at each scale forming scale-free topologies, but the specific links in existence vary dramatically between observation time scales as well as over time.

In summary, we have demonstrated that the static topology does not capture the dynamics of social networks. The prominence of nodes (as measured by degree) within the networks fluctuates widely from day to day, and a high degree in the aggregate network does not predict a high degree for individual days. Our conclusions are in sharp contrast to previous complex network research, which emphasizes the importance of aggregate nodal centrality in a static network topology [4-11,15-18].

Implications of a dynamic node centrality contrast with existing analyses that consider targeting nodes with the highest degrees to disrupt network communication or transport. Dynamic centrality implies that targeting nodes with the highest degrees at one time only weakly affects the nodes that are highly connected at another time. The approach of targeting high-degree nodes has been suggested, for example, to be an effective disease and computer virus prevention strategy; i.e. identification and "vaccination" of those nodes, would inhibit the spread of infection or computer viruses [10,18-20,22,23]. Our work implies that, at the very least, a more agile strategy of monitoring and vaccinating nodes based upon centrality over time is necessary. Otherwise a treatment based upon aggregate connectivity information will miss the impact of a node that otherwise has a low connectivity, becoming highly connected.

The type of dynamic analysis of networks we performed is pertinent to a wide range of network types. Whether or not there exists an underlying fixed topological structure, the question of which links are actually used is a relevant one. Thus, actual travel on a transportation network, and actual interactions that occur between molecules that can bind to each other, are both examples of networks that have an underlying structure but whose dynamic structure is relevant to the behavior of the system over time. In addition to the e-mail network studied here, we have found similar results when analyzing social network data about interactions found from the spatial proximity of personal Bluetooth wireless devices [21].

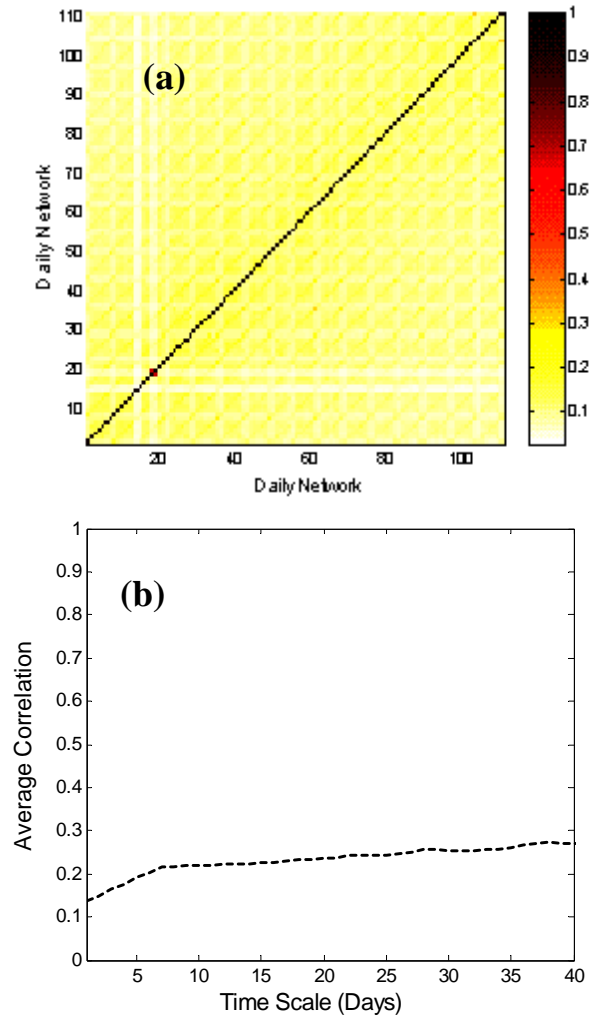

FIG. 1. **(a)** Matrix of correlations between pairs of daily networks sampled July 29[th], 2001 (Sunday) to November 18[th], 2001 (Sunday). Days 55 and 56 were excluded from further analysis due to lack of email communication. **(b)** Correlation between pairs of daily networks aggregated over times ranging from 1 to 40 days.

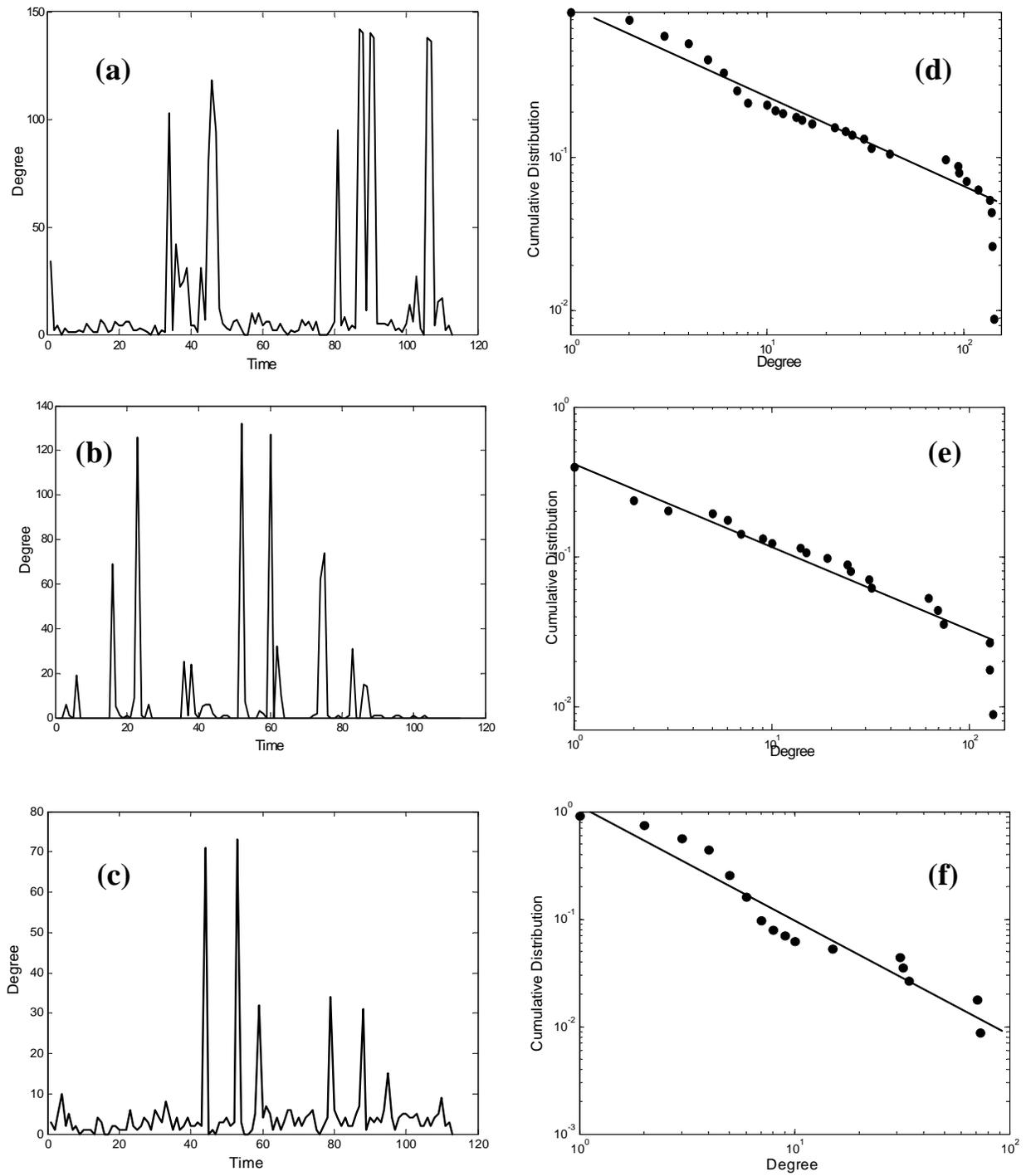

FIG. 2. Degree variations over time associated with the most connected node ('local hub') identified for a particular daily network. **(a-c)** Time series of degrees associated with nodes 724 (hub in day 34), 4631 (hub in day 52), and 450 (hub in day 44), respectively. Small and very large node degrees are observed. **(d-f)** The corresponding log-log plots of the cumulative distributions of degrees over time associated with 'local hubs' 724, 4631, and 450, respectively. The distributions follow a power law ( $p < 0.001$ ).

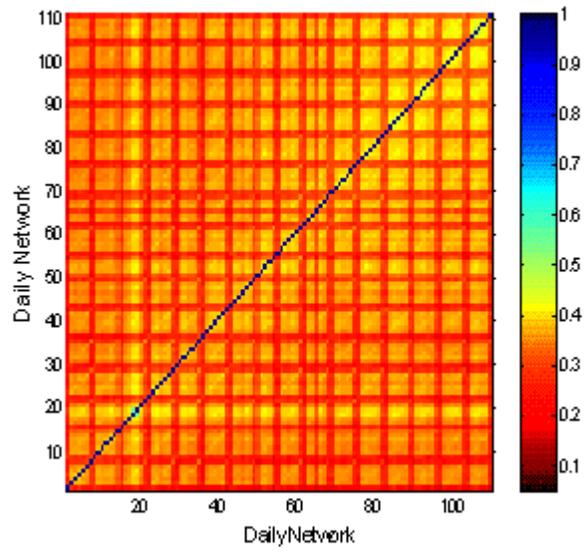

FIG. 3. Top-ranking list overlap between pairs of daily networks. For each pair of networks, the color code of the matrix denotes the percentage of nodes that appear in the 1000 top-ranking list of the networks.

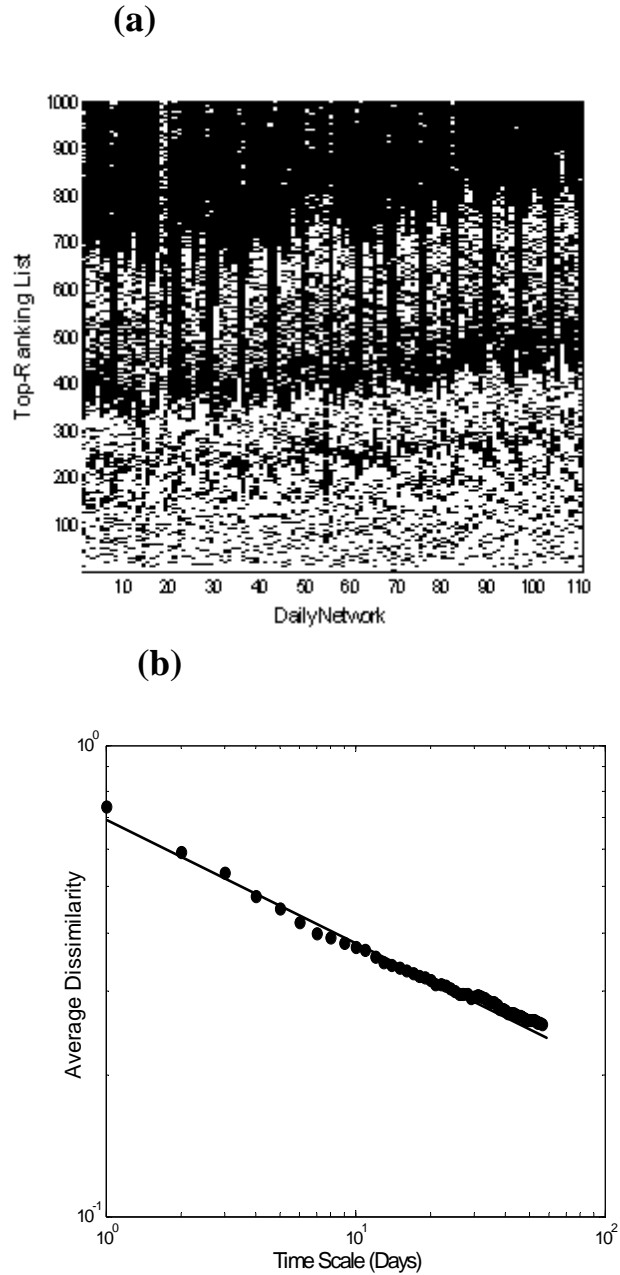

FIG. 4. **(a)** Comparison of the aggregate network with daily networks. A binary overlap matrix describing whether a node, included in the 1000 top-ranking list of a daily network, also appear (colored white) in the 1000 top-ranking list of the aggregate network. **(b)** Average dissimilarity of networks aggregated over times ranging from 1 to 56 days. Dissimilarity is measured as one minus the fractional overlap of the 1000 top-ranking nodes. The plot follows a power law ($p < 0.001$) indicating that networks formed over longer time periods do not converge to a well-defined structure.

11